\documentstyle[floats,aps]{revtex}
\parskip=\medskipamount
\begin{document}
\draft
\def\be{\begin{equation}}
\def\ee{\end{equation}}
\def\bea{\begin{eqnarray}}
\def\eea{\end{eqnarray}}
\def\cS{ {\cal{S}}  }
\def\cN{ {\cal{N}}  }
\def\cH{ {\cal{H}}  }
\def\cO{ {\cal{O}}  }
\def\cF{ {\cal{F}}  }
\def\bx{ {\bf x}  }
\def\e{ \epsilon  }
\def\nd{{ \vphantom{\nodagger}  } }
\def\ij{{ \langle ij\rangle  }}
\def\bA{ {\bf A}  }
\def\nd{ {\vphantom{\dagger}}  }

\twocolumn[\hsize\textwidth\columnwidth\hsize\csname @twocolumnfalse\endcsname

\title{Even-odd correlations in capacitance fluctuations of quantum dots}
\author{O. Prus ,  A. Auerbach,  Y. Aloni, U. Sivan}
\address{Solid State Institute and  Physics Department, Technion-IIT, Haifa 32000, Israel,}
\author{ and   R. Berkovits, } 
\address{The Jack and Pearl Resnick Inst. of Advanced Technology,\\ Dept. of Physics,
Bar Ilan University, Ramat Gan 52900, Israel.}
\thanks{Email: assa@pharaoh.technion.ac.il}
\date{\today}
\maketitle 

\begin{abstract} 
We investigate effects of short range interactions on the addition spectra of
quantum dots using a disordered Hubbard model.   A correlation function $\cS(q)$ is defined on the inverse compressibility  versus filling data, and computed numerically for small lattices. 
Two regimes of interaction strength are identified: the even/odd
fluctuations regime typical of Fermi liquid  ground states, and a regime of 
structureless $\cS(q)$ at strong interactions. We propose to understand the latter regime 
in terms of magnetically correlated localized spins.

\end{abstract}
\vskip2pc] 

\narrowtext
Coulomb interactions and  disorder in
electronic systems have posed a major
challenge to condensed matter physics for quite some time.  

Quantum dots with discrete electronic spectra offer a new
avenue to this problem.   A direct probe to the ground state energy is given by Coulomb blockade peaks in the conductance as the gate voltage is varied\cite{exp0,exp1,exp2,exp3}.  Theory of spectral fluctuations of non interacting electrons has made much progress
during the last decade due to the advent of  semiclassical approximations, random matrix theory and the non linear sigma model approach \cite{exp0,sing-el-theory}.   However  since Coulomb interactions are essential for the ``Coulomb blockade''  effect,  one may wonder
as to the validity of non interacting approximations to quantum dots in general. In particular: Is the  ground state qualitatively similar or different than 
a Fock state of the lowest single electron orbitals?

To gain insight into this question, we consider a  system of interacting electrons on a finite tight binding lattice with onsite
disorder.   The inverse compressibility at consecutive fillings is
\bea
\Delta(N) =    E(N+1) -2 E(N)+E(N-1) 
\label{Delta}
\eea
where $E(N)$ is the ground state energy of a dot with $N$ electrons.  (We assume weakly coupled leads such that $N$ is well defined within the area of the dot.)
By varying  a gate potential 
$\varphi$,
the dot's energy is modified to $E^\varphi_N = E(N) -e\varphi N$.  
Conductance peaks through the leads are observed   at
$E^\varphi(N)=E^\varphi(N+1) $, i.e. at potentials
$e\varphi_N=  E(N+1) - E(N)$. Thus differences between the peak potentials  $\varphi_N$ yield   direct measurements of $\Delta(N)$ which can be defined as  $e^{2}$ times  the {\em discrete inverse capacitance} of the dot.

We shall model the 
single electron part  of the dot's Hamiltonian by a site-disordered tight binding model
\bea
\cH_0&=&\sum_{is} w_i  c^\dagger_{is} c^\nd_{is}-\sum_{\ij} t_{ij}(B) c^\dagger_{is} c^\nd_{js}\\
&=& \sum_{n s} \e_n  \alpha^\dagger_{ns}\alpha^\nd_{ns} 
\label{h0}
\eea
where $c^\dagger_{is}$ creates an electron at site $i$ with spin $s$, $\ij$ denote nearest neighbors on the lattice, and $w_i $ are random site energies taken from a uniform distribution in the domain $[-W/\sqrt{3},W/\sqrt{3} ]$.  $\alpha^\dagger_{ns}$ creates an electron in eigenstate $\phi_n$ and spin $s$.  
An orbital coupling to a magnetic field is included by defining  $t_{ij}(B)= te^{ie\bA\cdot \bx_{ij}}$, where
$\nabla\times\bA=B$.

In the absence of electron interactions,  the inverse compressibility is given by 
\be
\Delta_0(N)   = \cases{ 0& $N=2n+1$\cr
\epsilon_{n+1} - \epsilon_n& $N=2n$}
\label{non-int}
\ee
where $\e_n$ are defined in (\ref{h0}). We find it useful to define the
``Ising''  variables
\be
S(N) \equiv   {\Delta(N) - \Delta(N-1)   \over \left|  \Delta(N)    - \Delta(N-1)\right|}
\ee
and a corresponding correlation function on a series of
$L$ consecutive data points,
\be 
\cS(q) \equiv {1\over L^2} \sum_{i,j=1}^{L} S(N_i) S(N_j) 
\exp\left(-i (N_i-N_j) q\right) 
\label{SQ}
\ee 
Obviously,  the non interacting   spectrum has perfect ``long range antiferromagnetic correlations''  i.e.  $\cS_0(\pi) = 1$. 

Coulomb interactions are treated by separating the interactions into the long and short range parts. A crude approximation to these two terms is given  by an  onsite and an infinite range  parts
\be
H_{int} = e^2{N(N-1)\over 2C} +U\sum_i n_{i\uparrow}n_{i\downarrow}.
\label{hint}
\ee
where $n_{is}=c^\dagger_{is}c_{is}, s=\uparrow,\downarrow$.
It is clear that the first term simply adds a constant  ${e^2\over C}$ to  $\Delta(N)$, and therefore does not alter $\cS(q)$. Thus in our model, deviations of $S(q)$ from $\cS_0(q)$ must therefore be a  consequence of the Hubbard  interactions described by the second term in (\ref{hint}).

We restrict ourselves to a  square lattice of  $\cN$  sites with  periodic boundary conditions, and to  disorder strength $W$  appropriate for the ``diffusive'' regime, i.e. the 
mean free path $l$ for the non interacting electrons  is of the order of, or smaller than, the system's linear length $L$. Using $l=v_F \tau$, where the inverse lifetime
is calculated in the Born approximation to be $\tau={8t\hbar    \over 2 \pi W^2}$,
we find $L/l={\pi\over 16 }\sqrt{\cN} (W/t)^2$.  

{\em Perturbation theory:} We  diagonalize $\cH_0$ on a square lattice of size $\cN$, with periodic boundary conditions  for a given realization  $\{w_i\}$. The single electron eigenenergies $\{\epsilon_n\}$ and wavefunctions $\psi_n(i)$  are assumed to be known.
The first order correction to Eq. (\ref{non-int}) are given by second differences of the  first order energies
\be
E_1(N) = U\sum_{n_\uparrow\le n_{F\uparrow}} \sum_{n_\downarrow\le n_{F\downarrow}} \left|\psi_{n_\uparrow}\right|^2  \left|\psi_{n_\downarrow}\right|^2 
\label{E1}
\ee
Here we appeal to the random matrix properties of 
$\cH_0$ in the diffusive regime,  in order to estimate
the magnitude and fluctuations of $E_1$ analytically.
We assume a  Random Vector Model (RVM) 
where all eigenvectors $\psi_n$
are random complex unit vectors  of dimension $N$ whose ensemble averaged correlations are\cite{mehta}  
\be
\langle \psi_n(\bx_i)\psi_m(\bx_j)\rangle_{rvm}  \propto \delta_{ij}\delta_{nm}
\ee
Using  the
orthonormalization  constraints for $\psi_n$ we obtain  after some algebra\cite{oleg-thesis} that the RVM estimate for the average first order correction to $\Delta$ is
\be
\Delta_1^{rvm}(N) =\cases{ {3U \over \cN+2}& $N$ odd\cr
-{2U \over \cN+2}& $N$ even}
\label{rvm}\ee
We have compared the RVM estimates to numerical results for disorder averaged $\Delta^{num}_1(N)$  for odd and even $N$ respectively.  Calculations have been done for
lattice sizes $\cN=56,110,210,420$, with the disorder varied in the range  $l/L\in[0.1,2.5]$. We find that
\be
\Delta^{num}_1(N) \in \cases{ [2.3,3.5] \times{U \over \cN}& $N$ odd\cr
[-1.3,-2.7]\times {U \over \cN}& $N$ even}
\label{num}\ee
Comparison of (\ref{rvm}) to (\ref{num}) shows the  RVM estimates 
to be in the right ball park as those gotten by numerically determining $\psi_n$  of the disordered tight binding model.
The main lesson learned by this calculation is that first order Hubbard corrections  {\em reduce} on average the fluctuations in $\Delta(N)$ since they are positive for odd $N$ and negative (on average) for even $N$. 

{\em Variational Theory for Sign Flips.}
The weak coupling regime is defined where the non interacting ground state $|\Psi_0\rangle$
is variationally stable against particle hole excitations.  When interaction strength  exceeds a certain threshold, it is variationally advantageous to create spin polarized electron-hole pairs e.g. $c^\dagger_{n_{F\uparrow}+1}c^\nd_{n_{F\downarrow}} |\Psi_0\rangle $
which reduce the Hubbard interaction energy at the expense of enhanced
single particle (kinetic) energy $\e_{n_F+1}-\e_{n_F}$.
In this variational theory,  for an even number of electrons the threshold for forming a triplet  is given by the inequality
\bea
0&>&\Delta\epsilon  -\left( E_1^t(2n)-E_1^s(2n) \right) 
=\Delta\epsilon   +U  \cF\nonumber\\
\cF&\equiv& \sum_i  \Bigg( \left(|\psi_{n+1} (\bx_i)|^2-|\psi_{n} (\bx_i)|^2\right) \left(\sum_{n'}^{n-1} |\psi_{n'} (\bx_i)|^2\right) 
 \nonumber\\
&&~~~~~~~~~~~~~~+|\psi_{n} (\bx_i)|^4\Bigg)
\label{ineq}
\eea
where $\Delta\epsilon=\epsilon_{n+1}-\epsilon_n$ and 
$E_1^t,E_1^s$ are the interaction corrections to the  singlet and triplet energies respectively.  Eq. (\ref{ineq})
can alternatively be  written as an inequality for  $\Delta(N)$ which includes up to first order corrections in $U$:
\bea
0 &>& \Delta(2n) -\Delta(2n-1)  \nonumber\\
&&~~~~-U \sum_i   \left(|\psi_{n+1} (\bx_i)|^2-|\psi_{n} (\bx_i)|^2\right) |\psi_{n} (\bx_i)|^2 
\label{trip-sing}
\eea
For extended random wave functions, the last term in (\ref{trip-sing}) is readily seen to be of order $ \cN^{-3/2}$ and thus  negligible in comparison to $\Delta(2n) -\Delta(2n-1)$.  
Eqs. (\ref{ineq},\ref{trip-sing}) establish the
connection between formation of  triplets  and sign flips of $S(N)$ (defined in (\ref{SQ})). The sign flips  degrade the 
``antiferromagnetic'' correlations of $\cS(q)$ as $U$ is increased. We shall proceed to estimate the leading dependence of $\cS(\pi;U)$ by a statistical calculation based on the properties of the non interacting spectrum of $\cH_0$  with and without an orbital magnetic field. 

We assume  that the level spacing statistics of $H_0$ in the diffusive regime of $L/l\simeq 1$ is that of a random matrix in the Gaussian Orthogonal Ensemble  (GOE).   
In the presence of
a magnetic field $B$ whose flux through the dot is of order of one flux quantum divided by $\cN^{1/4}$\cite{AS} ,
the level  spacing statistics of $\cH_0$ turns into
the Gaussian Unitary Ensemble (GUE). These assumptions were
checked numerically and verified quite well for the Hamiltonian (\ref{h0}) with and without external flux, on lattices up to $\cN=420$  sites.

The Wigner  distributions of $\Delta\e$ are\cite{mehta}  
\bea
P_2^{GOE} (\Delta\epsilon ) &=&  {\Delta\epsilon\pi \over 2 {\bar \Delta}^2 } e^{-{\pi(\Delta\epsilon)^2\over 4  {\bar \Delta}^2 }  } \nonumber\\
P_2^{GUE} (\Delta\epsilon ) &=&  {32(\Delta\epsilon)^2 \over \pi^2 {\bar \Delta}^3 } e^{-{4 (\Delta\epsilon)^2\over \pi {\bar \Delta}^2 }  } 
\eea
where ${\bar\Delta}$ is the mean level spacing.

The probablity distribution of the interaction  term  in (\ref{ineq}) is denoted by
 \be
P_1(U\cF)= {1\over U}P_1'(\cF) ,
\ee
where $P_1'$ is a dimensionless function of its dimensionless argument.  The probability of satisfying Eq. (\ref{ineq}) is given 
by the double integral
\bea
P^{flip}(U)&=& \int_0^\infty d\cF P_1'(\cF)\int_0^{U\cF} d(\Delta\e) P_2(\Delta\e)
\eea
which at weak coupling $U << {\bar\Delta}$ we can evaulate using the low energy
expansion of $P_2$ and obtain
\bea
P^{flip}_{GOE}&\approx  U^2 {\pi\over 2{\bar\Delta}^2}\int_0^\infty dx P_1'(x) \int_0^x d\e \e = &U^2 {\pi\over 4 {\bar\Delta}^2}\langle \cF^2\rangle\nonumber\\
P^{flip}_{GUE}&\approx  U^3 {32\over\pi^2  {\bar\Delta}^3}\int_0^\infty dx P_1'(x) \int_0^x d\e \e = &U^2 {32\over 3\pi^2 {\bar\Delta}^3}\langle \cF^3\rangle
\label{pflip}
\eea
At weak coupling the reduction in the perfect even-odd correlations are proportional to
$P^{flip}$, and therefore we find that
\be
1-\cS(\pi)  \propto P^{flip}(U) \propto \cases{U^2& GOE\cr
U^3&GUE}
\label{predict}\ee

{\em Numerical Diagonalizations}. The variational theory and the random matrix estimates were checked by numerically
diagonalizing the full interacting Hamiltonian on a lattice size $3 \times 3$ with periodic boundary conditions with  a magnetic flux $\phi$
threading the lattice at its center. We varied the number of electrons for each
specific realization from $N=5$ to $N=13$, each time diagonalizing
the corresponding $\left(_{2 \cN}^{N}\right)$ matrix (i.e., maximum matrices
of size $48620 \times 48620$ for $N=9$) for various values
of $U$. Then $S(\pi)$ was calculated and averaged over 500 different
realizations.

Two different values of $\phi$ were considered:
$\phi=0$ and $\phi= \pi/2$ which should correspond
to the GOE and GUE statistics. 
Because of the
small size of the samples
the maximum flux of $\phi= \pi/2$ is not strong enough to completely remove
time reversl symetry, and the level spacing is a combination of the GOE and
GUE level spacing. The results of $S(\pi)$ for both cases
are presented in Fig. \ref{fig1}. It can be seen that the general
behavior predicted in  Eqs. (\ref{predict}) is observed. A reasonable fit for  the numerical results in the regime $U \le 1$ is obtained by  $S(\pi) = 1 - 0.3U^2$  for
$\phi=0$  and $S(\pi) = 1 - 0.2U^3$. This supports the validity of the GOE
(GUE) statistics for the case $\phi=0$  ($\phi=\pi/2$).

{\em Experimental $\cS(q)$}.
In Fig. (\ref{fig2}) we have plotted three sets of data for $\Delta(N)$ and $\cS(q)$. The top two sets were measured on two GaAs samples at different fillings.  The bottom data, measured on InO,  was taken from Ref.\cite{IO}.  It is interesting to note that $\cS(q)$ shows strong even-odd correlations only for the middle data set, while
it has no remnance of these correlations for the top and bottom sets. Although we do not attempt to provide a quantitative understanding of the data in this paper, we dare to
speculate that the flatness of $\cS(q)$ correlations is definitevely associated with  strongly correlated samples
where ground states  significantly differ from Fermi-liquid  Fock states.
The fluctuations in $\Delta(N)$ due to the direct Coulomb interaction and its interplay with disorder are discussed in Ref. \cite{ew}.

{\em Acknowledgements}
We acknowledge useful discussions with the late Arcadi Aronov, 
and thank Yuval Gefen for helpful comments.
The support of US-Israeli Binational Science Foundation, the Israeli Academy
of Sciences and the Fund for Promotion of Research at Technion are gratefully acknowledged.

\vfill\eject 
\begin{figure}
\caption{Averaged $\protect \cS(q)$, Eq. ({\protect\ref{SQ}}),  for the disordered Hubbard model on a $3 \times 3$ torus. Inset:  ${\protect \cS(\pi)}$ for the ${\protect\phi=0}$ and ${\protect\phi=\pi/2}$
cases,  fitted to estimates of RVM variational theory.}
\label{fig1}
\end{figure}
\begin{figure}
\caption{Experimental second energy differences from quantum dots of GaAs, Ref.{\protect\cite{Aloni}} (top and middle),
and InO, Ref.{\protect\cite{IO}}, (bottom), with corresponding  correlations $\cS(q)$.
Correlations indicates that the top and bottom datae are in the strongly interacting regime.}
\label{fig2}
\end{figure}
\end{document}